\documentclass[twocolumn,twoside]{IEEEtran}
\usepackage{mathpazo}
\usepackage{times}
\usepackage{color}
\usepackage{amsmath}
\usepackage{amsfonts}
\usepackage{latexsym}
\usepackage{amssymb}
\usepackage{cite}

\usepackage{upref}
\usepackage{theorem}
\usepackage{graphicx}
\usepackage{psfrag}

\theoremstyle{plain}
\theorembodyfont{\normalfont\slshape}

\newtheorem{thm}{Theorem$\!$}
\newenvironment{theorem}
{\begin{thm}\hspace*{-1ex}{\bf.}}{\end{thm}}

\newtheorem{lem}[thm]{Lemma$\!$}
\newenvironment{lemma}{\begin{lem}\hspace*{-1ex}{\bf.}}{\end{lem}}

\newtheorem{prop}[thm]{Proposition$\!$}

\newtheorem{cor}[thm]{Corollary$\!$}

\newtheorem{defn}[thm]{Definition$\!$}
\newenvironment{definition}{\begin{defn}\hspace*{-1ex}{\bf.}}{\end{defn}}

\newtheorem{xmpl}[thm]{Example$\!$}
\newenvironment{example}{\begin{xmpl}\hspace*{-1ex}{\bf.}}{\hfill$\Box$\end{xmpl}}

\newtheorem{cnstr}{Construction$\!$}

\newenvironment{construction}{\begin{cnstr}\hspace*{-1ex}{\bf.}}{\hfill$\Box$\end{cnstr}}

\newcounter{enumrom}
\renewcommand{\theenumrom}{(\roman{enumrom})}

\makeatletter
\renewcommand{\@endtheorem}{\endtrivlist}
\makeatother

\makeatletter
\renewcommand{\thefigure}{{\@arabic\c@figure}}
\renewcommand{\fnum@figure}{{\bf Figure\,\thefigure}}
\makeatother

\newcommand{\cC}{\mathcal{C}}
\newcommand{\cD}{\mathcal{D}}
\newcommand{\cE}{\mathcal{E}}

\newcommand{\cG}{\mathcal{G}}

\newcommand{\cL}{\mathcal{L}}
\newcommand{\cM}{\mathcal{M}}

\newcommand{\cP}{\mathcal{P}}

\newcommand{\mathset}[1]{\left\{#1\right\}}
\newcommand{\abs}[1]{\left|#1\right|}

\newcommand{\floorenv}[1]{\left\lfloor #1 \right\rfloor}
\newcommand{\parenv}[1]{\left( #1 \right)}
\newcommand{\sparenv}[1]{\left[ #1 \right]}

\newcommand{\be}[1]{\begin{equation}\label{#1}}
\newcommand{\ee}{\end{equation}}

\renewcommand{\leq}{\leqslant}
\renewcommand{\ge}{\geqslant}
\renewcommand{\geq}{\geqslant}

\newcommand{\Cref}[1]{Co\-ro\-lla\-ry\,\ref{#1}}

\newcommand{\gf}{{\mathrm{GF}}}

\newcommand{\sbinom}[2]{\genfrac{[}{]}{0pt}{}{#1}{#2}}
\newcommand{\spn}[1]{\left\langle {#1} \right\rangle}

\outer\def\proclaim #1. #2\par{\medbreak
 \noindent{\bf#1.\enspace}{\sl#2\par}%
 \ifdim\lastskip<\medskipamount \removelastskip\penalty55\medskip\fi}

\begin{document}


\title{\Huge\bf Gray Codes and Enumerative Coding \\ for Vector Spaces}

\author{\large
Moshe~Schwartz,~\IEEEmembership{Senior Member,~IEEE}
\thanks{This work was supported in part by ISF grant 134/10.}
\thanks{Moshe Schwartz is with the Department
   of Electrical and Computer Engineering, Ben-Gurion University of the Negev,
   Beer Sheva 8410501, Israel
   (e-mail: schwartz@ee.bgu.ac.il).}
}

\maketitle

\begin{abstract}
Gray codes for vector spaces are considered in two graphs: the
Grassmann graph, and the projective-space graph, both of which have
recently found applications in network coding. For the Grassmann
graph, constructions of cyclic optimal codes are given for all
parameters. As for the projective-space graph, two constructions for
specific parameters are provided, as well some non-existence results.

Furthermore, encoding and decoding algorithms are given for the
Grassmannian Gray code, which induce an enumerative-coding scheme. The
computational complexity of the algorithms is at least as low as known
schemes, and for certain parameter ranges, the new scheme outperforms
previously-known ones.
\end{abstract}


\section{Introduction}

\IEEEPARstart{G}{ray} codes, named after their inventor, Frank Gray
\cite{Gra53}, were originally defined as a listing of all the binary
words, each appearing exactly once, such that adjacent words in the
list differ by the value of a single bit. Since then, numerous
generalizations were made, where today, a Gray code usually means a
listing of the elements of some space, such that each element appears
no more than once, and adjacent elements are ``similar''. What
constitutes similarity usually depends on the application of the code.

The use of Gray codes has reached a wide variety of areas, such as
storage and retrieval applications \cite{ChaCheChe92}, processor
allocation \cite{CheShi90}, statistics \cite{DiaHol94}, hashing
\cite{Fal88}, puzzles \cite{Gar72}, ordering documents \cite{Los92},
signal encoding \cite{Lud81}, data compression \cite{Ric86}, circuit
testing \cite{RobCoh81}, measurement devices \cite{SchEtz99}, and
recently also modulation schemes for flash memories
\cite{JiaMatSchBru09,EngLanSchBru11b,YehSch12}. For a survey on Gray
codes the reader is referred to \cite{Sav97}.

In the past few years, interest has grown in $q$-analogs of
combinatorial structures, in which vectors and subsets are replaced by
vector spaces over a finite field. Two prominent examples are the
Grassmann graph $\cG_q(n,k)$, and the projective-space graph
$\cP_q(n)$. The former contains all the $k$-dimensional subspaces of
an $n$-dimensional vector space over $\gf(q)$, and is the $q$-analog
of the Johnson graph, whereas the latter contains all the subspaces of
an $n$-dimensional vector space, and acts as the $q$-analog of the
Hamming graph.

Examples of such $q$-analogs structures are codes and anti-codes in
the Grassmann graph \cite{FraWil86,SchEtz02}, Steiner systems
\cite{BraEtzOstVarWas13}, reconstruction problems
\cite{YaaSchLanBru13}, and the middle-levels problem \cite{Etz13}. But
what has begun as a purely theoretical area of research, has recently
found an important application to network coding, starting with the
work of Koetter and Kschischang \cite{KoeKsc08}, and continuing with
\cite{XiaFu09,EtzVar11,SilKscKoe08,SilKsc09,GadYan10a,GadYan10b,EtzSil09,Ska10}.

In this work we study $q$-analogs of Gray codes, which are Hamiltonian
circuits in the projective-space graph, and $q$-analogs for
constant-weight Gray codes, which are Hamiltonian circuits in the
Grassmann graph. For the former, we present non-existence results
(both for cyclic and non-cyclic codes), as well as constructions for
specific parameters based on the middle-levels problem discussed in
\cite{Etz13}. For the latter, we provide constructions for cyclic
optimal Gray codes for all parameters, as well as encoding and
decoding functions. The construction has many degrees of freedom,
resulting a large number of Gray codes, which we bound from below.

As a side effect of the Gray-code construction and the encoding and
decoding algorithms we provide, we obtain an enumerative-coding scheme
for the Grassmannian space. A general enumerative-coding algorithm due
to Cover \cite{Cov73}, was recently used as the basis for an
enumerative-coding scheme specifically designed for the Grassmannian
space $\cG_q(n,k)$ by Silberstein and Etzion \cite{SilEtz11}, who
provided encoding and decoding algorithms with complexity $O(M[nk]n)$,
where $M[m]$ denotes the number of operations required for multiplying
two numbers with $m$ digits each. Another work by Medvedeva
\cite{Med12} suggested only a decoding algorithm with complexity
$O(M[n^2]\log n)$. We provide encoding and decoding algorithm that not
only arrange the subspaces in a Gray code, but also operate in
$O(M[nk]n)$ time, the same complexity as the algorithms of
\cite{SilEtz11}.  We provide another decoding algorithm of complexity
$O(M[nk]k\log k)$, which outperforms the decoding algorithm of
\cite{SilEtz11} when $k\log k=o(n)$ (for example, when
$k=O(n^{1-\epsilon})$ for some $0<\epsilon < 1$), and outperforms the
decoding algorithm of \cite{Med12} when $k=o(\sqrt{n})$.

The paper is organized as follows: In Section \ref{sec:prem} we
provide the basic definitions and notation used throughout the paper.
In Section \ref{sec:grass} we construct Grassmannian Gray codes, as
well as provide encoding and decoding functions. We continue in
Section \ref{sec:proj} by studying subspace Gray codes. We conclude in
Section \ref{sec:conc} with a summary and open problems.

\section{Preliminaries}
\label{sec:prem}

Throughout the paper we shall maintain a notation consisting of
upper-case letters for vector spaces, sometimes with a superscript
indicating the dimension. We shall denote vectors by lower-case
letters, and scalars by Greek letters.  For a vector space $W$ over
some finite field $\gf(q)$, we let $\dim(W)$ denote the dimension of
$W$. For two subspaces, $W_1$ and $W_2$, $W_1+W_2$ will denote their
sum.  If that sum happens to be a direct sum, we'll stress that fact
by denoting it as $W_1\oplus W_2$. For a vector $v\in\gf(q)^n$, we
shall denote the space spanned by $v$ as $\spn{v}$.

Let $W^n$ be some fixed $n$-dimensional vector space over
$\gf(q)$. For an integer $0\leq k\leq n$, we denote by
$\sbinom{W^n}{k}$ the set of all $k$-dimensional subspaces of $W^n$.

\begin{definition}
The Grassmann graph $\cG_q(n,k)=(V,E)$ is defined by the vertex set
$V=\sbinom{W^n}{k}$, and two vertices $W_1,W_2\in V$ are connected by an
edge iff $\dim(W_1\cap W_2)=k-1$.
\end{definition}

The
$q$-number of $k$ is defined as
\[[k]_q=1+q+q^2+\dots+q^{k-1}=\frac{q^k-1}{q-1}.\]
By abuse of notation we denote
\[[k]_q!=[k]_q[k-1]_q\dots [1]_q.\]

The \emph{Gaussian coefficient} is defined for $n$, $k$, and $q$
as
\begin{align*}
\sbinom{n}{k}_q&=\frac{[n]_q!}{[k]_q![n-k]_q!} \\
&=\frac{(q^n-1)(q^{n-1}-1)\dots(q^{n-k+1}-1)}{(q^k-1)(q^{k-1}-1)\dots(q-1)}.
\end{align*}
It is well known that the number of $k$-dimensional subspaces of an
$n$-dimensional space over $\gf(q)$ is given by $\sbinom{n}{k}_q$. Furthermore,
the Gaussian coefficients satisfy the following recursion
\[\sbinom{n}{k}_q=\sbinom{n-1}{k}_q + q^{n-k}\sbinom{n-1}{k-1}_q,\]
as well as the symmetry
\[\sbinom{n}{k}_q=\sbinom{n}{n-k}_q,\]
for all integers $0\leq k\leq n$ (for example, see \cite{LinWil01}).

Another graph of interest is the following.
\begin{definition}
The \emph{projective-space graph} $\cP_q(n)=(V,E)$ is defined by the
vertex set $V=\bigcup_{k=0}^n \sbinom{W^n}{k}$, and two vertices
$W_1,W_2\in V$ are connected by an edge iff 
\[\dim(W_1)+\dim(W_2)-2\dim(W_1\cap W_2)=1.\]
\end{definition}

Equivalently, two vertices, $W_1$ and $W_2$, are connected in
$\cP_q(n)$ iff $\abs{\dim(W_1)-\dim(W_2)}=1$, and either $W_1\subset
W_2$ or $W_2\subset W_1$.

We now provide the definitions for the Gray codes that we study in
this paper.

\begin{definition}
Let $W^n$ be an $n$-dimensional vector space over $\gf(q)$. An
\emph{$(n,k;q)$-Grassmannian Gray code}, $\cC$, is a sequence of distinct
subspaces
\[\cC=C_0,C_1,\dots,C_{P-1},\]
where $C_i\in\sbinom{W^n}{k}$, and where $C_i$ and $C_{i+1}$ are
neighbors in $\cG_q(n,k)$, for all $0\leq i\leq P-2$. We say $P$ is
the size of the code $\cC$. If $C_0$ and $C_{P-1}$ are neighbors in
$\cG_q(n,k)$ then $\cC$ is said to be \emph{cyclic} and $P$ is its
\emph{period}. If $P=\sbinom{n}{k}_q$, then $\cC$ is called
\emph{optimal}.
\end{definition}

A similar definition holds for the graph $\cP_q(n)$.
\begin{definition}
Let $W^n$ be an $n$-dimensional vector space over $\gf(q)$. An
\emph{$(n;q)$-subspace Gray code}, $\cC$, is a sequence of distinct
subspaces
\[\cC=C_0,C_1,\dots,C_{P-1},\]
where $C_i\in\bigcup_{k=0}^n\sbinom{W^n}{k}$, and where $C_i$ and
$C_{i+1}$ are neighbors in $\cP_q(n)$ for all $0\leq i\leq P-2$. We
say $P$ is the size of the code $\cC$. If $C_0$ and $C_{P-1}$ are
neighbors in $\cP_q(n)$ then $\cC$ is said to be \emph{cyclic} and $P$
is its \emph{period}. If $P=\sum_{k=0}^n\sbinom{n}{k}_q$, then $\cC$
is called \emph{optimal}.
\end{definition}

\section{Grassmannian Gray Codes}
\label{sec:grass}

In this section we will study Grassmannian Gray codes. We will first
describe a construction, and later introduce and analyze encoding and
decoding algorithms. These algorithms may be used as an
enumerative-coding scheme.

\subsection{Construction}

The construction we describe is recursive in nature. We will be
constructing an $(n,k;q)$-Grassmannian Gray code by combining together
an $(n-1,k;q)$-code with an $(n-1,k-1;q)$-code. We start by
introducing two useful lemmas.

\begin{lemma}
\label{lem:add1}
Let $W^n$ be an $n$-dimensional vector space over $\gf(q)$. Let
$W^{n-1}$ and $W^{k-1}$ be $(n-1)$-dimensional and $(k-1)$-dimensional
subspaces of $W^n$, respectively, where 
\[W^{k-1}\subseteq W^{n-1}\subset W^n.\]
Then there are $q^{n-k}$ vectors $v_0,\dots,v_{q^{n-k}-1}\in\gf(q)^n$ such
that:
\begin{enumerate}
\item
  $\parenv{W^{k-1}\oplus\spn{v_i}}\cap W^{n-1}=W^{k-1}$.
\item
  The subspaces $W^{k-1}\oplus\spn{v_i}$ are distinct
\end{enumerate}
\end{lemma}
\begin{IEEEproof}
To maintain the first requirement it is obvious that $v_i\in
W^n\setminus W^{n-1}$. We have
\[\abs{W^n\setminus W^{n-1}}=q^{n}-q^{n-1}\]
vectors to choose from. However, having chosen a vector $v\in
W^n\setminus W^{n-1}$, to maintain the first requirement we cannot
choose vectors of the form $\alpha v + w$, where $w\in W^{k-1}$ and
$\alpha\in\gf(q)\setminus\mathset{0}$. Since there are $(q-1)q^{k-1}$
distinct choices of $\alpha$ and $w$, resulting in distinct forbidden
vectors, the maximal number of vectors we can choose which maintain
the two requirements is given by
\[\frac{q^n-q^{n-1}}{(q-1)q^{k-1}}=q^{n-k}.\]
\end{IEEEproof}

A closer look at the proof of Lemma \ref{lem:add1} reveals that
$W^{k-1}$ induces an equivalence relation on the vectors of
$W^n\setminus W^{n-1}$, where $v,v'\in W^n\setminus W^{n-1}$ are
equivalent if there exist $\alpha\in\gf(q)\setminus\mathset{0}$ and
$w\in W^{k-1}$ such that $v'=\alpha v+w$. A set of vectors whose
existence is guaranteed by Lemma \ref{lem:add1} is merely a list of
representatives from each of the equivalence classes induced by
$W^{k-1}$. For such a vector $v$ and a subspace $W^{k-1}$, we shall
denote the equivalence class of $v$ induced by $W^{k-1}$ as
$[v]_{W^{k-1}}$.

\begin{lemma}
\label{lem:distincteq}
Let $W^{n-1}$ and $W^n$, be as in Lemma \ref{lem:add1}. Assume $W^{k-1}_1$ and
$W^{k-1}_2$ are two distinct $(k-1)$-dimensional subspaces of $W^{n-1}$. Then
for any $v\in W^n\setminus W^{n-1}$ we have
\[ [v]_{W^{k-1}_1} \neq [v]_{W^{k-1}_2}.\]
\end{lemma}
\begin{IEEEproof}
We observe that
\[\spn{[v]_{W^{k-1}_1}}=\spn{v}\oplus W^{k-1}_1 \quad \text{and}
\spn{[v]_{W^{k-1}_2}}=\spn{v}\oplus W^{k-1}_2.\]
Let us assume to the contrary that
\[ [v]_{W^{k-1}_1} = [v]_{W^{k-1}_2}.\]
We therefore have
\begin{align*}
W^{k-1}_1 &= W^{n-1}\cap\parenv{\spn{v}\oplus W^{k-1}_1}\\
& =  W^{n-1} \cap \spn{[v]_{W^{k-1}_1}}\\
& =  W^{n-1} \cap \spn{[v]_{W^{k-1}_2}}\\
& = W^{n-1}\cap\parenv{\spn{v}\oplus W^{k-1}_2} \\
& = W^{k-1}_2,
\end{align*}
which is a contradiction.
\end{IEEEproof}

Intuitively speaking, Lemma \ref{lem:distincteq} states that the
equivalence classes that partition $W^{n}\setminus W^{n-1}$ and are
induced by distinct $(k-1)$-dimensional subspaces of $W^{n-1}$, do not
contain two identical classes. This fact will be used later in the
construction.

We shall now build an $(n,k;q)$-Grassmannian Gray code by combining an
$(n-1,k;q)$-code with an $(n-1,k-1;q)$-code.

\begin{construction}
\label{con:grass}
Let $W^n$ be an $n$-dimensional vector space over $\gf(q)$. We can
write $W^n$ as the direct sum $W^n=W^{n-1}\oplus W^1$, where
$\dim(W^{n-1})=n-1$ and $\dim(W^1)=1$.

Let us assume the existence of two cyclic optimal Grassmannian Gray
codes: an $(n-1,k;q)$-code $\cC'$, and an $(n-1,k-1;q)$-code $\cC''$. In
both cases we assume the ambient vector space is $W^{n-1}$. For
convenience, let us denote the code sequences as
\begin{align*}
\cC' &= C'_0,C'_1,\dots,C'_{P'-1}, \\
\cC'' &= C''_0,C''_1,\dots,C''_{P''-1}.
\end{align*}
From these two codes we shall construct a new $(n,k;q)$-Grassmannian
Gray code.

We start with $C''_0$, and choose equivalence-class representatives
$v''_{0,0},\dots,v''_{0,q^{n-k}-1}$ by Lemma
\ref{lem:add1}. Continuing to $C''_1$, again we choose
equivalence-class representatives,
$v''_{1,0},\dots,v''_{1,q^{n-k}-1}$, where we make sure
\[ [v''_{0,q^{n-k}-1}]_{C''_0}\cap [v''_{1,0}]_{C''_1}\neq \emptyset,\]
i.e., that the last equivalence class chosen for $C''_0$, and the
first equivalence class chosen for $C''_1$, have a non-empty
intersection.

We continue in the same manner, where for $C''_i$ we choose equivalence-class
representatives $v''_{i,0},\dots,v''_{i,q^{n-k}-1}$, where also
\[ [v''_{i-1,q^{n-k}-1}]_{C''_{i-1}}\cap [v''_{i,0}]_{C''_i}\neq \emptyset.\]
Finally, for $C''_{P''-1}$, the last subspace in $C'$, we need both a
non-empty intersection of
\[ [v''_{P''-2,q^{n-k}-1}]_{C''_{P''-2}}\cap [v''_{P''-1,0}]_{C''_{P''-1}}\neq \emptyset,\]
as well as a non-empty intersection of
\[ [v''_{P''-1,q^{n-k}-1}]_{C''_{P''-1}}\cap [v''_{0,0}]_{C''_0}\neq \emptyset,\]
i.e., with the first equivalence class induced by the first subspace
$C_0$.  Since, by Lemma \ref{lem:distincteq}, $[v''_{0,0}]_{C''_0}$
has a non-empty intersection with at least two equivalence classes
induced by $C''_{P''-1}$, we can always find a suitable set of
representatives.

We now construct the auxiliary sequence $\cC^*$ as follows:
\begin{align*}
\cC^* = &C''_0\oplus\spn{v''_{0,0}},C''_0\oplus\spn{v''_{0,1}},
\dots,C''_0\oplus\spn{v''_{0,{q^{n-k}-1}}},\\
&C''_1\oplus\spn{v''_{1,0}},C''_1\oplus\spn{v''_{1,1}},
\dots,C''_1\oplus\spn{v''_{1,{q^{n-k}-1}}},\\
&\vdots \\
&C''_{P''-1}\oplus\spn{v''_{P''-1,0}},
\dots,C''_{P''-1}\oplus\spn{v''_{P''-1,{q^{n-k}-1}}}.
\end{align*}
In a more concise form, 
\[C^*=C^*_0,C^*_1,\dots,C^*_{P''q^{n-k}-1}\]
is a sequence of length $P''q^{n-k}$ in which the $i$th element is the
subspace
\[\cC^*_i=C''_{\floorenv{i/q^{n-k}}}\oplus\spn{v''_{\floorenv{i/q^{n-k}},i \bmod q^{n-k}}}.\]

We now turn to use the code $\cC'$. Let us choose an arbitrary index
$0\leq j\leq P'-1$, and denote $U=C'_j\cap C'_{j+1}$, where the
indices are taken modulo $P'$. We observe that $U\subseteq W^{n-1}$ is
a $(k-1)$-dimensional subspace.

Since $\cC'$ contains all the $(k-1)$-dimensional subspaces of
$W^{n-1}$, let $i$ be the index such that $C''_i=U$. Finally, we also choose
an arbitrary index $0\leq \ell \leq q^{n-k}-2$.

We now construct the code $\cC$ by inserting a shifted version of
$\cC'$ into the auxiliary $\cC^*$ as follows:
\begin{align}
\cC=&C^*_0,C^*_1,\dots,C^*_{iq^{n-k}+\ell} \nonumber\\
&C'_{j+1},C'_{j+2},\dots,C'_{P'-1},C'_0,C'_1,\dots,C'_j\nonumber\\
&C^*_{iq^{n-k}+\ell+1},\dots,C^*_{P''q^{n-k}-1}.
\label{eq:insert}
\end{align}
\end{construction}

\begin{theorem}
The sequence $\cC$ of subspaces from Construction \ref{con:grass} is a
cyclic optimal $(n,k;q)$-Grassmannian Gray code.
\end{theorem}
\begin{IEEEproof}
We start by showing that the subspaces in the code are all
distinct. We first note that the subspaces in $\cC^*$ are distinct
from those in $\cC'$, since all the former intersect $W^{n-1}$ in a
$(k-1)$-dimensional subspace, while all the latter intersect $W^{n-1}$
in a $k$-dimensional subspace. To continue, the subspaces of $\cC'$
are distinct by virtue of $\cC'$ being a Grassmannian Gray
code. Finally, we show that the subspaces of $\cC^*$ are
distinct. Assume
\[C''_{i_1}\oplus\spn{v''_{i_1,j_1}}=C''_{i_2}\oplus\spn{v''_{i_2,j_2}}.\]
Then
\begin{align*}
C''_{i_1}&=\parenv{C''_{i_1}\oplus\spn{v''_{i_1,j_1}}}\cap W^{n-1}\\
&=\parenv{C''_{i_2}\oplus\spn{v''_{i_2,j_2}}}\cap W^{n-1}=C''_{i_2}.
\end{align*}
Since $\cC''$ is a Grassmannian Gray code, we must have $i_1=i_2$. We thus have
\[C''_{i_1}\oplus\spn{v''_{i_1,j_1}}=C''_{i_1}\oplus\spn{v''_{i_1,j_2}}.\]
Since the vectors $v''_{i_1,0},\dots,v''_{i_1,q^{n-k}-1}$ were chosen
from distinct equivalence classes, we again must have
$j_1=j_2$. Hence, all the subspaces of $\cC$ are distinct.

Next, we show that any two subspaces which adjacent in the list,
intersect in a $(k-1)$-dimensional subspace. This is certainly true
for adjacent subspaces in $\cC'$ since they form an
$(n-1,k;q)$-Grassmannian Gray code. For $\cC^*$ we have
\[\parenv{C''_i\oplus\spn{v''_{i,j}}} \cap
\parenv{C''_i\oplus\spn{v''_{i,j+1}}}=C''_i\] and so the intersection
is $(k-1)$-dimensional. Furthermore, $C''_i$ and $C''_{i+1}$ intersect
in a $(k-2)$-dimensional subspace, since they come from a
$(n-1,k-1;q)$-Grassmannian Gray code. Since, by construction,
\[ [v''_{i,q^{n-k}-1}]_{C''_i}\cap [v''_{i+1,0}]_{C''_{i+1}} \neq \emptyset,\]
we have
\[ \dim\parenv{\parenv{C''_{i}\oplus\spn{v''_{i,q^{n-k}-1}}}
\cap\parenv{C''_{i+1}\oplus\spn{v''_{i+1,0}}}} = k-1.\]
Let $i$, $j$,
and $\ell$, be as in \eqref{eq:insert}.  We can also easily verify
that at the insertion points of $\cC'$ into $\cC^*$ we have
\begin{align*}
\dim\parenv{C^*_{iq^{n-k}+\ell}\cap C'_{j+1}} & =k-1, \\
\dim\parenv{C^*_{iq^{n-k}+\ell+1}\cap C'_{j}} & =k-1,
\end{align*}
and thus, all adjacent subspaces in the sequence are also adjacent in the
graph $\cG_q(n,k)$. This also proves the code is cyclic.

Finally, to show that the code is optimal we need to show that it
contains all the $k$-dimensional subspaces of $W^n$. Since $\cC'$ and
$\cC''$ are optimal we have
\begin{align*}
\abs{\cC}&=\abs{\cC'}+q^{n-k}\abs{\cC''}\\
&=\sbinom{n-1}{k}_q+q^{n-k}\sbinom{n-1}{k-1}_q=\sbinom{n}{k}_q
=\abs{W^{n}}.
\end{align*}
\end{IEEEproof}

\begin{theorem}
For every $n\geq 1$ and $0\leq k\leq n$ there exists a cyclic optimal
$(n,k;q)$-Grassmannian Gray code.
\end{theorem}
\begin{IEEEproof}
Because of the recursive nature of Construction \ref{con:grass}, the
only thing we need to prove is that the basis for the recursion
exists. This is trivially true since $(n,n;q)$-Grassmannian Gray codes
and $(n,0;q)$-Grassmannian Gray codes which are cyclic and optimal are
the unique sequence of length $1$ containing the full vector space,
and the trivial space of dimension $0$, respectively.
\end{IEEEproof}

We can get a lower bound on the number of distinct
$(n,k;q)$-Grassmannian Gray codes that result from this construction,
thus getting a lower bound on the number of such codes in general. The
counting requires the following lemma.

\begin{lemma}
\label{lem:intersect}
Let $W^{n-1}$ and $W^{n-1}$ be as in lemma \ref{lem:add1}, and let
$C_1,C_2\subset W^{n-1}$ be two $(k-1)$-dimensional subspaces such
that $\dim(C_1\cap C_2)=k-2$. Then for any $v_1\in W^n\setminus
W^{n-1}$, there exist exactly $q$ distinct subspaces of the form
$C_2\oplus\spn{v_2}$, for some $v_2\in W^n\setminus W^{n-1}$, such
that
\[\dim\parenv{\parenv{C_1\oplus\spn{v_1}}\cap\parenv{C_2\oplus\spn{v_2}}}=k-1.\]
\end{lemma}
\begin{IEEEproof}
Let $w_1,w_2,\dots,w_{k-2}$ be a basis for $C_1\cap C_2$. Let us further denote
\begin{align*}
C_1 &= \spn{w_1,\dots,w_{k-2},u_1}, \\
C_2 &= \spn{w_1,\dots,w_{k-2},u_2}.
\end{align*}
Given $v_1\in W^n\setminus W^{n-1}$, in order to obtain a subspace
$W_2\oplus\spn{v_2}$ with the desired intersection dimension we must
choose $v_2\in W^n\setminus W^{n-1}$ such that the equation
\[ \sum_{i=1}^{k-2} \alpha_i w_i + \alpha u_1 + \beta v_1=
\sum_{i=1}^{k-2} \gamma_i w_i + \gamma u_2 + \delta v_2,\] holds for
some choice of scalar coefficients
$\alpha_i,\gamma_i,\alpha,\beta,\gamma,\delta\in\gf(q)$, with
$\beta\neq 0$. We thus choose
\[v_2 = \frac{1}{\delta}\sum_{i=1}^{k-2} (\alpha_i-\gamma_i) w_i + \alpha u_1 -\gamma u_2 + \beta v_1.\]

Since multiplying $v_2$ by a scalar does not change the subspace
$C_2\oplus\spn{v_2}$, we may conveniently choose $\delta=\beta^{-1}$. Hence,
\[v_2 = \sum_{i=1}^{k-2} \frac{\alpha_i-\gamma_i}{\beta} w_i + \frac{\alpha}{\beta} u_1 -\frac{\gamma}{\beta} u_2 + v_1.\]
Finally, we note that adding a vector from $C_2$ to $v$ does not change
the subspace $C_2\oplus\spn{v_2}$. We may therefore eliminate any linear combination of $w_1,\dots,w_{k-2},u_2$ from $v$. By denoting $\epsilon=\alpha/\beta$,
we are left with choosing
\[v_2=v_1+\epsilon u_1,\]
and there are exactly $q$ choices for $\epsilon\in\gf(q)$ which result
in distinct subspaces as required.
\end{IEEEproof}

We are now ready to state the lower bound on the number of distinct
$(n,k;q)$-Grassmannian Gray codes resulting from Construction \ref{con:grass}.
We note that codes which are cyclic shifts of one another are still counted
as distinct codes.

\begin{theorem}
The number of distinct $(n,k;q)$-Grassmannian Gray codes resulting
from Construction \ref{con:grass} is lower bounded by
\[\prod_{i=1}^{n-k}\prod_{j=1}^{k}
\parenv{ (q-1)q^{i-1} \parenv{\parenv{q^i-1}!q}^{\sbinom{i+j-1}{j-1}_q}}^{\binom{n-i-j}{n-k-i}}.\]
\end{theorem}
\begin{IEEEproof}
Let us denote the number of $(n,k;q)$-Grassmannian Gray codes by
$T(n,k;q)$.  If either $k=n$ or $k=0$, then $T(n,k;q)=1$, which agrees
with the claimed lower bound. Let us therefore consider the case of
$0<k<n$.

During the construction process, we first choose an
$(n-1,k-1;q)$-code, which can be done in $T(n-1,k-1;q)$ ways. We then
need to choose the vectors $v''_{i,j}$ to obtain the subspaces
$C''_i\oplus\spn{v''_{i,j}}$. For $i=0$ we can arrange the $q^{n-k}$
subspaces in $\parenv{q^{n-k}}!$ ways. For subsequent values of $i$,
$1\leq i\leq \sbinom{n-1}{k-1}_q-2$, we can choose the first subspace
$C''_i\oplus\spn{v''_{i,0}}$ in one of $q$ ways, according to Lemma
\ref{lem:intersect}. The rest of the subspaces may be chosen
arbitrarily in any one of $\parenv{q^{n-k}-1}!$ ways. Finally, for
$i=\sbinom{n-1}{k-1}_q-1$, both the first subspace and last subspace
are chosen from a set of $q$ subspaces. At the worst case, we can
choose them both in one of $q(q-1)$ ways, and the rest of the
subspaces in $\parenv{q^{n-k}-2}!$.

We then choose an $(n-1,k;q)$-code, which can be done in $T(n-1,k;q)$
ways. We rotate and insert it into the code constructed so
far. However, since we already count cyclic shifts of codes as
distinct, we shall assume we do not rotate it, to avoid
over-counting. We, thus, only have to choose where to insert it, in
one of $q^{n-k}-1$ ways.

Combining all of the above, we reach the recursion,
\begin{align*}
T(n,k;q) &\geq T(n-1,k-1;q)T(n-1,k;q) \cdot\\
&\qquad \cdot (q-1)q^{n-k-1}\parenv{\parenv{q^{n-k}-1}!q}^{\sbinom{n-1}{k-1}_q}.
\end{align*}
Solving the recursion, with the base cases of $T(n,n;q)=1$ and
$T(n,0;q)=1$, gives the desired lower bound.
\end{IEEEproof}

\subsection{Algorithms}

We now describe algorithms related to Grassmannian Gray codes. The
algorithms we consider are:
\begin{enumerate}
\item
Encoding -- Finding the $i$th element in the code.
\item
Decoding -- Finding the index in the list of a given element of the code.
\end{enumerate} 
We will specialize Construction \ref{con:grass} to allow for simpler
algorithms.

We require some more notation. Throughout this section we denote by
$e_i$ the $i$th standard unit vector, i.e., the vector all of whose
entries are $0$ except for the $i$th one being $1$. The length of the
vector will be implied by the context. The entries of a length $n$
vector will be indexed by $0,1,\dots,n-1$.  The $n\times n$ identity
matrix will be denoted by $I_n$, and the $n_1\times n_2$ all-zero
matrix by $0_{n_1\times n_2}$.

A $k$-dimensional subspace $W^k$ of an $n$-dimensional space can be
represented by a $k\times n$ matrix whose rows form a basis for $W^k$.
Many choices for such a matrix exist, and we shall be interested in a
unique one. We will first describe the reduced row echelon form
matrix, which is known to be unique, and then transform it to obtain
our representation.

In a reduced row echelon form matrix, the leading coefficient of each
row is $1$, and it is the only non-zero element in its
column. Furthermore, the leading coefficient of each row is strictly
to the right of the leading coefficient of the previous row.

Assume $M$ is a $k\times n$ matrix of rank $k$ in reduced row echelon
form, $k\leq n$. We denote the set of $k$ indices of columns
containing leading coefficients as
$\Lambda(M)\subseteq\mathset{0,1,\dots,n-1}$.  We apply the following
simple recursive transformation $\tau$ to $M$: If $k=0$ then $\tau(M)$
is the degenerate empty matrix with $0$ rows. Otherwise, assume $k\geq
1$. If the last column of $M$ is all zeros, then
$\tau(M)=[\tau(M^*)|0_{k\times 1}]$, where $M^*$ is the $k\times
(n-1)$ matrix obtained from $M$ by deleting the last column. If the
last column of $M$ is not all zeros, let $i$ be the index of the first
row from the bottom which does not contain a zero in the last
column. We multiply the $i$th row by a scalar such that its last entry
is $1$. We then subtract suitable scalar multiples of the $i$th row
from other rows of $M$ so that the resulting matrix $M'$ has a single
non-zero entry in the last column (a $1$ located in the $i$th row). We
then delete the $i$th row and the last column to get the
$(k-1)\times(n-1)$ matrix $M''$. We recursively take $\tau(M'')$,
append a column of $0$'s to its right, and re-insert the $i$th column
which we previously removed. The result is defined as $\tau(M)$.

\begin{example}
Let $M$ be the $3\times 5$ reduced row echelon form matrix
\[M=\begin{pmatrix}
1 & 0 & 3 & 0 & 1 \\
0 & 1 & 2 & 0 & 4 \\
0 & 0 & 0 & 1 & 2
\end{pmatrix},
\]
where the entries are from $\gf(5)$. We then have
\[\tau(M)=\begin{pmatrix}
1 & 1 & 0 & 0 & 0 \\
0 & 2 & 4 & 1 & 0 \\
0 & 0 & 0 & 3 & 1
\end{pmatrix}
\]
\end{example}

It is easily seen that $\tau(M)$ is in row echelon form, but not in
\emph{reduced} row echelon form, i.e., the leading coefficient of each
row is non-zero (but not necessarily $1$), the entries below a leading
coefficient are $0$ (but not necessarily $0$ above it), and the
leading coefficient of each row is strictly to the right of the
leading coefficient of the previous row. We note that
$\Lambda(M)=\Lambda(\tau(M))$.

Thus, for a $k$-dimensional subspace $W^k$, and the unique reduced row
echelon form matrix $M$ whose rows form a basis for $W^k$, we shall
call $\tau(M)$ the \emph{canonical} matrix representation of $W^k$.
To avoid excessive notation we shall refer to both the subspace and
its canonical matrix as $W^k$. We say $W^k$ is \emph{simple} if
\[W^{k}=\sparenv{ I_k | 0_{n-k}}.\]

We now start with specializing Construction \ref{con:grass}. First,
during the construction we require a choice of $W^n$ and $W^{n-1}$. We
choose both to be simple subspaces.

Next, in the construction we have $\cC''=C''_0,\dots,C''_{P''-1}$, and
for each $C''_i$, a $(k-1)$-dimensional subspace of $W^{n-1}$, we find
$q^{n-k}$ vectors from $W^n\setminus W^{n-1}$, denoted
$v''_{i,0},\dots,v''_{i,q^{n-k}-1}$. We make this choice explicit: let
$C''_i$ be a $(k-1)\times (n-1)$ canonical matrix. Let $0\leq
r_{i,0}<r_{i,1}<\dots < r_{i,n-k-1}\leq n-2$ be the elements in
$\mathset{0,1,\dots,n-2}\setminus\Lambda(C''_i)$. We note that
$e_{r_{i,\ell}}$ is not in the subspace $C''_i$, for all $\ell$. For
an integer $0\leq j\leq q^{n-k}-1$, let $[j]_\ell$ denote its $\ell$th
digit when written in base $q$, i.e.,
\[j=\sum_{\ell=0}^{n-k-1}[j]_\ell q^\ell,\]
where $[j]_\ell\in\mathset{0,1,\dots,q-1}$. For convenience, we also
denote the elements of $\gf(q)$ as
$\alpha_0,\alpha_1,\dots,\alpha_{q-1}$, in some fixed order, where
$\rho(\cdot)$ gives the reverse mapping, i.e., $\rho(\alpha_i)=i$. We
now choose
\[v''_{i,j} = e_{n-1} + \sum_{\ell=0}^{n-k-1} \alpha_{[j]_\ell} e_{r_{i,\ell}}.\]

In Construction \ref{con:grass}, when inserting a shifted version of
$\cC'$, a parameter $0\leq \ell \leq q^{n-k}-2$ is chosen. We shall
call this parameter the \emph{insertion offset}. In this instance, we
will always choose $\ell=0$.

Finally, we say a cyclic optimal $(n,k;q)$-Grassmannian Gray code,
$\cC=C_0,C_1,\dots,C_{P-1}$, is \emph{simple}, if $C_0$ is simple, and
$C_0\cap C_{P-1}$ is simple.

\begin{lemma}
Let $\cC'$ be a simple cyclic optimal $(n-1,k;q)$-Grassmannian Gray
code, and $\cC''$ be a simple cyclic optimal
$(n-1,k-1;q)$-Grassmannian Gray code. Let $\cC=C_0,C_1,\dots,C_{P-1}$
be the cyclic optimal $(n,k;q)$-Grassmannian Gray code created by
Construction \ref{con:grass}, with an insertion offset $\ell=0$. Then
its shifted version,
\[\overline{\cC} = C_1,C_2,\dots,C_{P-1},C_0,\]
is a simple cyclic optimal $(n,k;q)$-Grassmannian Gray code.
\end{lemma}
\begin{IEEEproof}
The fact that $\overline{\cC}$ is a cyclic optimal $(n,k;q)$-code is
trivial. It remains to prove the code is simple. Let us denote
\begin{align*}
\cC' &= C'_0,C'_1,\dots,C'_{P'-1}, \\
\cC'' &= C''_0,C''_1,\dots,C''_{P''-1}.
\end{align*}
Since $\cC'$ is simple, we have that $C'_0$ is simple, and that
$C'_0\cap C'_{P'-1}$ is simple. The latter intersection determines
where $\cC'$ is inserted in $\cC^*$, i.e., between the $q^{n-k}$
subspaces derived from the simple $(k-1)$-dimensional space. Since
$\cC''$ is also simple, it is inserted in the first set of $q^{n-k}$
subspaces derived from $C''_0$.  By using an insertion offset
$\ell=0$, we have that $C_1=C'_0$, and that $C_1\cap C_0=C''_0$. Thus,
$\overline{\cC}$ is simple.
\end{IEEEproof}

We are now in a position to state simple encoding and decoding
functions.  The encoding function
\[\cE_{n,k;q}:\mathset{0,1,\dots,\sbinom{n}{k}_q-1}\to \sbinom{W^n}{k},\]
maps an index $m$ to the the $m$th subspace in the
$(n,k;q)$-Grassmannian Gray code $\overline{\cC}$ constructed
above. Using the observations so far, we can easily state that
\[
\cE_{n,k;q}(m)=\begin{cases}
[I_k|0_{k\times(n-k)}] & \text{$k=0$ or $k=n$,} \\
[\cE_{n-1,k;q}(m)|0_{k\times 1}] & m\leq \sbinom{n-1}{k}_q-1,\\
\begin{bmatrix}
\cE_{n-1,k-1;q}(i)|0_{(k-1)\times 1} \\
v''_{i,j}
\end{bmatrix} & \text{otherwise,}
\end{cases}
\]
where
\begin{align*}
i & = \floorenv{\frac{1}{q^{n-k}}\parenv{m-\sbinom{n-1}{k}_q+1}}
\bmod \sbinom{n-1}{k-1}_q\\
j & = \parenv{m-\sbinom{n-1}{k}_q+1} \bmod q^{n-k} \\
v''_{i,j} & = e_{n-1} + \sum_{\ell=0}^{n-k-1} \alpha_{[j]_\ell} e_{r_{i,\ell}}.
\end{align*}
We also note, that in the last case, where the vector $v''_{i,j}$ is
appended as another row to the generating matrix, the vector is
inserted between the correct rows such that the resulting matrix is
canonical.

The decoding function,
\[\cD_{n,k;q}:\sbinom{W^n}{k}\to \mathset{0,1,\dots,\sbinom{n}{k}_q-1},\]
is defined as the reverse of the encoding function, i.e.,
\[\cD_{n,k;q}\parenv{\cE_{n,k;q}(m)}=m,\]
for all $0\leq m\leq \sbinom{n}{k}_q-1$.

To describe the decoding function we need some preparation
work. Assume the input to the decoding function is a $k$-dimensional
subspace $W^k$, which will also denote a canonical matrix whose row
span is $W^k$.

Since $W^{n-1}$ is simple, checking whether $W^k\subseteq W^{n-1}$
amounts to checking whether the last column of $W^k$ contains only
$0$'s. If this is not the case, then $\dim(W^k\cap W^{n-1})=k-1$, and
by construction there is a unique row with a non-zero entry in the
last coordinate. We denote this row $v''$, and its last coordinate
must be $1$. Furthermore, we remove this row and the last column from
$W^k$ and denote the resulting canonical matrix by $W^{k-1}$. Finally,
like before, let $0\leq r_{0}<r_{1}<\dots < r_{n-k-1}\leq n-2$ be
elements of $\mathset{0,1,\dots,n-2}\setminus\Lambda(W^{k-1})$.

With this notation we can now easily find that
\begin{equation}
\label{eq:decode}
\cD_{n,k;q}(W^{k}) = \begin{cases}
0 & \text{$k=0$ or $k=n$,}\\
\cD_{n-1,k;q}(W^{k}) & W^{k}\subseteq W^{n-1}, \\
i & \text{otherwise,}
\end{cases}
\end{equation}
where
\begin{multline*}
i=\sbinom{n-1}{k}_q+
\Bigg(
\Bigg(
q^{n-k}\cdot\cD_{n-1,k-1;q}(W^{k-1})-1 \\
+\sum_{\ell=0}^{n-k-1}\rho\parenv{e_{r_\ell}\cdot v''}q^{\ell}
\Bigg)
\bmod q^{n-k}\sbinom{n-1}{k-1}_q
\Bigg).
\end{multline*}
We also note that in the case of $W^{k}\subseteq W^{n-1}$, when
applying $\cD_{n-1,k;q}$ to $W^{k}$ we remove the last column of
$W^{k}$, which is an all-zero column.

\subsection{Complexity Analysis}

The goal of this section is to bound the number of operations required
to perform the encoding and decoding procedures described in the
previous section.

For our convenience, we assume throughout this section that integers
are represented in base $q$. Thus, multiplying and dividing by $q$
amount to simple shift operations on the list of digits.

Another simplification is enabled by the following lemma.
\begin{lemma}
\label{lem:dual}
Let $\cC=C_0,C_1,\dots,C_{P-1}$ be an $(n,k;q)$-Grassmannian Gray code.
Then the dual code,
\[\cC^\perp=C_0^\perp,C_1^\perp,\dots,C_{P-1}^\perp,\]
is an $(n,n-k;q)$-Grassmannian Gray code. If $\cC$ is cyclic, then so
is $\cC^\perp$. Also, if $\cC$ is optimal, then so is $\cC^\perp$.
\end{lemma}
\begin{IEEEproof}
Obviously $\dim(C_i^\perp)=n-k$ for all $i$. Since $(C_i^\perp)^\perp=C_i$,
the elements of $\cC^\perp$ are all distinct. To verify that adjacent
elements in $\cC^\perp$ are also adjacent in $\cG_q(n,n-k)$ we use
simple linear algebra. For all $i$,
\[C_i^\perp \cap C_{i+1}^\perp=\parenv{C_i+C_{i+1}}^\perp.\]
Since $\dim(C_i\cap C_{i+1})=k-1$, we have
\begin{align*}
\dim(C_i+C_{i+1})&=\dim(C_i)+\dim(C_{i+1})\\
&\quad\ -\dim(C_i\cap C_{i+1})\\
&=k+1.
\end{align*}
It then follows that
\[\dim(C_i^\perp \cap C_{i+1}^\perp) = 
n-\dim\parenv{C_i+C_{i+1}}^\perp = n-k-1,\]
hence, $C_i^\perp$ and $C_{i+1}^\perp$ are adjacent in $\cG_q(n,n-k)$.
If we take all indices modulo $P$, then $\cC^\perp$ is cyclic if $\cC$ is
cyclic. Finally, $\sbinom{n}{k}_q=\sbinom{n}{n-k}_q$ implies that $\cC^\perp$
is optimal if $\cC$ is optimal.
\end{IEEEproof}

In light of Lemma \ref{lem:dual}, we will assume throughout that
$2k\leq n$, and in particular, that $n-k=\Theta(n)$.

An important ingredient in the analysis is the complexity of
multiplying two numbers, each with $m$ digits. We denote this number
as $M[m]$. Using the Sch\"onhage-Strassen algorithm we have
$M[m]=O(m\log m \log\log m)$ (for example, see \cite{Knu97v2}). We can
alternatively use the more recent algorithm due to F\"urer
\cite{Fur09}, for which $M[m]=O(m\log m 2^{\log^* m})$. We also note
that division of two numbers with $m$ digits each also requires
$O(M[m])$ operations \cite{Knu97v2}.

We now turn to the analysis of the decoding algorithm. We observe that
all the integers involved require at most $nk$ digits to represent.
As a first
step, we compute $\sbinom{n}{k}_q$. It was shown in \cite{SilEtz11}
that the complexity of this is $O(M[nk]k)$\footnote{
To be more precise, it was shown in \cite{SilEtz11} that computing
$\sbinom{n-1}{k}_q$ takes $O(kn(n-k)\log n \log\log n)$ operations.
To facilitate
a comparison with the complexity analysis we perform, and by taking
$n-k=\Theta(n)$ due to Lemma \ref{lem:dual}, we may rewrite the result
of \cite{SilEtz11} as $O(M[nk]k)$. We will implicitly perform this
translation whenever comparing with \cite{SilEtz11}, and later, with
\cite{Med12}.
}.  As was also shown in
\cite{SilEtz11}, from this Gaussian coefficient we may derive
$\sbinom{n-1}{k-1}_q$ and $\sbinom{n-1}{k}_q$ by
\begin{align}
\label{eq:g1}
\sbinom{n}{k}_q &= \sbinom{n-1}{k-1}_q \cdot \frac{q^{n}-1}{q^{k}-1}, \\
\label{eq:g2}
\sbinom{n}{k}_q &= \sbinom{n-1}{k}_q \cdot \frac{q^{n}-1}{q^{n-k}-1},
\end{align}
with additional $O(M[nk])$
operations.

As we examine the algorithm as given in \eqref{eq:decode}, even though
it is presented as a recursive algorithm, it is a tail recursion, and
so, may be considered as an iterative process. At the beginning of
each iteration we check the last column of the matrix to see if it is
all $0$'s. This takes $O(k)$ time.

For the second case of \eqref{eq:decode}, when $W^k\subseteq W^{n-1}$,
we delete the last column, taking $O(k)$ time. For the third case of
\eqref{eq:decode}, we need to compute $\sbinom{n-1}{k}_q$ and
$\sbinom{n-1}{k-1}_q$ from $\sbinom{n}{k}_q$, taking $O(M[nk])$
operations. Multiplication by $q$ amounts to a simple shift operation,
and addition and subtraction of numbers with $nk$ digits takes $O(nk)$
time. We note that finding the numbers $r_\ell$ for $0\leq \ell\leq
n-k-1$ is easily seen to take at most $O(n)$ time. Deleting a row and
a column takes $O(n)$ time. Finally, we note that the sole purpose of
the modulo operation is to transform a possible $-1$ outcome into
$q^{n-k}\sbinom{n-1}{k-1}_q-1$ which may be done in $O(nk)$ time
(since we have already computed $\sbinom{n-1}{k-1}_q$). The total
number of operations for the last case of the decoding procedure is
therefore bounded by $O(M[nk])$. Since the total number of rounds is at most
$n$, the entire algorithm may be run in time $O(M[nk]n)$. The same analysis
holds for the encoding algorithm.

\begin{theorem}
The computation complexity of the encoding and decoding algorithms is
$O(M[nk]n)$.
\end{theorem}

The complexity of the algorithms from \cite{SilEtz11} is the same as
those presented in this work. However, the algorithms here also
provide a Gray ordering of the subspaces. We also mention
\cite{Med12}, in which only a decoding algorithm was suggested,
without Gray coding, achieving complexity of $O(M[n^2]\log n)$.

We can, however, improve the complexity of the decoding procedure for
a certain asymptotic range of $k$ by changing the way we compute
\eqref{eq:decode}. We start by changing the way we compute the
Gaussian coefficients. By definition,
\[\sbinom{n}{k}_q=\frac{(q^n-1)(q^{n-1}-1)\dots(q^{n-k+1}-1)}{(q^k-1)(q^{k-1}-1)\dots(q-1)}.\]
Our strategy to compute this value is to compute separately the
numerator and denominator, and then perform division. To compute the
numerator, we partition the $k$ parentheses into pairs and compute
their product, partition the $k/2$ results into pairs, and so on. For
ease of presentation we can assume $k$ is a power of $2$ to avoid
rounding, and this has no effect on the overall complexity.
Initially, each of the numbers in the numerator may be represented by
$n$ digits in base $q$. Thus, the total number of operations to
compute the numerator is
\[\sum_{i=1}^{\log k}\frac{k}{2^i}M[2^{i-1}n]=\frac{nk}{2}\sum_{i=1}^{\log k}\frac{M[2^{i-1}n]}{2^{i-1}n}\leq M[nk/2]\log k,\]
where the last inequality is due to the fact that $M[m]/m$ is a
non-decreasing function. The same analysis applies to the
denominator. Finally, we need to divide the numerator and denominator,
each with at most $nk$ digits, thus requiring additional $O(M[nk])$
operations. It follows that computing $\sbinom{n}{k}_q$ requires
$O(M[nk]\log k)$.

The analysis of the remaining part of the algorithm is nearly the
same. The only difference is that we do not use \eqref{eq:g1} and
\eqref{eq:g2} at every iteration. Instead, whenever we find ourselves
in the third case of \eqref{eq:decode} we compute the necessary
Gaussian coefficients from scratch.  We now make the crucial
observation that the algorithm takes at most $n$ iterations, at most
$k$ of which take the third case of \eqref{eq:decode}. Thus, the total
number of operations for a decoding procedure is $O(M[nk]k\log k)$.

\begin{theorem}
The decoding algorithm may be run using $O(M[nk]k\log k)$ operations.
\end{theorem}

We note that when $k\log k = o(n)$ (for example, when
$k=O(n^{1-\epsilon})$ for some $0<\epsilon < 1$) the decoding
algorithm we presented outperforms the decoding algorithm of
\cite{SilEtz11}, including the $O(n^2 k^2)$ decoding algorithm of
\cite{SilEtz11} for the smaller range of $k < \log n \log\log
n$. Furthermore, when $k=o(\sqrt{n})$, the decoding algorithm we
presented outperforms the decoding algorithm of \cite{Med12}.

\section{Subspace Gray Codes}
\label{sec:proj}

This section is devoted to study of subspace Gray codes. Unlike
optimal Grassmannian Gray codes, which exist for all parameters, the
case of subspace Gray codes appears to be more complicated. We begin
with nonexistence results, and then continue to constructing subspace
Gray code for a limited set of cases.

\subsection{Nonexistence Results}

The next two theorems show that for half of the parameter space,
optimal subspace Gray codes (cyclic or not) do not exist.

\begin{theorem}
\label{th:nosub}
There are no cyclic optimal $(n;q)$-subspace Gray codes when $n\geq 2$ is
even, over any finite field $\gf(q)$.
\end{theorem}
\begin{IEEEproof}
Let $n=2m$, $m\geq 1$. Assume to the contrary such a code $\cC$
exists, and $\cC=C_0,C_1,\dots,C_{P-1}$. By the definition of the
code, every time an $m$-dimensional subspace appears in the sequence,
it is followed (perhaps cyclically) by an $(m+1)$-dimensional subspace
or an $(m-1)$-dimensional subspace. Since the code is optimal, all
subspaces appear and so we must have
\[\sbinom{2m}{m+1}_q+\sbinom{2m}{m-1}_q \geq \sbinom{2m}{m}_q.\]
However,
\begin{equation}
\label{eq:ratio}
\frac{\sbinom{2m}{m}_q}{\sbinom{2m}{m+1}_q+\sbinom{2m}{m-1}_q} =
\frac{\sbinom{2m}{m}_q}{2\sbinom{2m}{m-1}_q}=\frac{1}{2}\cdot\frac{q^{m+1}-1}{q^{m}-1}> 1,
\end{equation}
for all $q\geq 2$.
\end{IEEEproof}
\begin{theorem}
There are no optimal $(n;q)$-subspace Gray codes (not necessarily cyclic)
when $n\geq 2$ is even, except for $n=2$ and $q=2$.
\end{theorem}
\begin{IEEEproof}
The proof is similar to that of Theorem \ref{th:nosub}. Again, let
$n=2m$, $m\geq 1$, and assume to the contrary such a code $\cC$
exists, and $\cC=C_0,C_1,\dots,C_{P-1}$. By the definition of the
code, every time an $m$-dimensional subspace appears in the sequence,
it is followed by an $(m+1)$-dimensional subspace or an
$(m-1)$-dimensional subspace, except if it is the last in the
sequence. Since the code is optimal, all subspaces appear and so we
must have
\begin{equation}
\label{eq:minusone}
\sbinom{2m}{m+1}_q+\sbinom{2m}{m-1}_q \geq \sbinom{2m}{m}_q-1.
\end{equation}
However, by the proof of Theorem \ref{th:nosub} we already know that
\[\sbinom{2m}{m+1}_q+\sbinom{2m}{m-1}_q < \sbinom{2m}{m}_q.\]
Thus, the only way for \eqref{eq:minusone} to hold is that
\[\sbinom{2m}{m+1}_q+\sbinom{2m}{m-1}_q = \sbinom{2m}{m}_q-1.\]
Using \eqref{eq:ratio}, we therefore need
\begin{equation}
\label{eq:need}
\sbinom{2m}{m}_q=1+\frac{2q^m-2}{q^{m+1}-2q^m+1}.
\end{equation}

When $q\geq 4$, and for all $m\geq 1$, we have
\[0 < 2q^m-2 < q^{m+1}-2q^m+1,\]
and so the RHS of \eqref{eq:need} is not an integer.

When $q=3$, for similar reasons, the RHS of \eqref{eq:need} is not an
integer except when $m=1$, but then
\[\sbinom{2}{1}_3=4\neq 2 = 1+ \frac{2\cdot 3^1-2}{3^2-2\cdot 3^1+1}.\]

Finally, when $q=2$, \eqref{eq:need} becomes
\[\sbinom{2m}{m}_2=2^{m+1}-1.\]
We observe that
\begin{align*}
\sbinom{2m}{m}_2&=\frac{[2m]_2!}{[m]_2![m]_2!}\\
&=\frac{(2^{2m}-1)(2^{2m-1}-1)\dots(2^{m+1}-1)}{(2^m-1)(2^{m-1}-1)\dots(2^1-1)}\\
&\ge2 \frac{2^{2m-1}\cdot 2^{2m-2} \cdot \dots \cdot 2^m}{2^m \cdot 2^{m-1}
\cdot \dots \cdot 2^1}\\
&= 2^{m(m-1)}.
\end{align*}
For $m\geq 3$ we have
\[2^{m(m-1)}> 2^{m+1}-1.\]
Thus, to complete the proof we only need to check the case of $m=2$,
for which we find that
\[\sbinom{4}{2}_2=35\neq 7 = 2^{3}-1.\]
\end{IEEEproof}

We note that there does indeed exist an optimal non-cyclic
$(2;2)$-subspace Gray code:
\[\cC=\spn{(1,0)},\spn{(0,0)},\spn{(0,1)},\spn{(0,1),(1,0)},\spn{(1,1)}.\]

\subsection{Constructions}

We now turn to the question of whether cyclic optimal $(n;q)$-subspace
Gray codes exist when $n$ is odd. The answer is trivial when $n=1$. We
also answer this in the positive for the cases of $n=3,5$ by using the
$q$-analog solution to the middle-level problem given in \cite{Etz13}.
We first describe the $q$-analog of the middle-level problem, and then
show how a solution there gives a cyclic optimal subspace Gray code.

Let $n=2m+1$ be an odd positive integer, and let $W^n$ be a vector
space over $\gf(q)$. We consider the following graph $\cM_q(2m+1)$:
the vertex set of the graph is $\sbinom{W^n}{m}\cup\sbinom{W^n}{m+1}$,
and two vertices $W_1$ and $W_2$ are connected by an edge iff
$W_1\subset W_2$ or $W_2\subset W_1$. An $(n;q)$-subspace Gray code
for the middle levels is a Hamiltonian path in $\cM_q(n)$, and it is
cyclic if it is a Hamiltonian circuit.

Etzion \cite{Etz13} proved the following theorem:
\begin{theorem}{\cite{Etz13}}
\label{th:etzion}
For any $q$, a power of a prime, there exists a cyclic optimal
$(3;q)$-subspace Gray code for the middle levels.
\end{theorem}

Using Theorem \ref{th:etzion}, we can prove the following theorem.
\begin{theorem}
For any $q$, a power of a prime, there exists a cyclic optimal
$(3;q)$-subspace Gray code.
\end{theorem}
\begin{IEEEproof}
Let $\cC'$ be the code guaranteed by Theorem \ref{th:etzion},
\[\cC'=C_0,C_1,\dots,C_{P'-1},\]
where
\[P'=\sbinom{3}{1}_q+\sbinom{3}{2}_q=2(q^2+q+1).\]
We note that $P'$ is even. Since this code contains all the subspaces
in the middle levels, the only two vertices of $\cP_q(3)$ not covered
are $W^3$, the entire space, and $W^0$, the $0$-dimensional trivial
subspace.

Since $\cC'$ is cyclic, let us assume, without loss of generality,
that $\dim(C_0)=1$. We now pick an arbitrary odd integer $1\leq i\leq
P'-3$, and construct the sequence,
\[\cC= W^0,C_0,C_1,\dots,C_i,W^3,C_{P'-1},C_{P'-2},\dots,C_{i+1}.\]

We contend $\cC$ is a cyclic optimal $(3;q)$-subspace Gray
code. Trivially, $\cC$ contains all the subspaces of $W^3$ exactly
once. Furthermore, since originally $\dim(C_i)=1$ iff $i$ is even, and
$\dim(C_i)=2$ iff $i$ is odd, the resulting sequence is indeed a
cyclic subspace Gray code.
\end{IEEEproof}

For the construction of $(5;q)$-subspace Gray codes we require a more
in-depth view of Etzion's construction from \cite{Etz13}. Let $W_1$
and $W_2$ be two $m$-dimensional subspaces of $W^n$ over some finite
field $\gf(q)$. We say $W_1$ and $W_2$ are equivalent if there exists
some $\alpha\in\gf(q)$ such that
\[W_1=\alpha W_2=\mathset{\alpha w ~|~ w\in W_2}.\]
It is easy to see that this is indeed an equivalence relation, and the
equivalence classes were called \emph{necklaces} in \cite{Etz13}. As
also noted in \cite{Etz13}, if $\gcd(n,m)=1$ then the size of any
equivalence class is $\sbinom{n}{1}_q$, and in particular, does not
depend of $m$.

Etzion proved the following two theorems, which will be the starting
point for our next construction.

\begin{theorem}{\cite{Etz13}}
\label{th:etzneck}
Let $n=2m+1$ and let $W^n$ be an $n$-dimensional vector space over
$\gf(q)$, with $\alpha\in\gf(q)$ a primitive element. Assume
$\cL=X_0,Y_0,X_1,Y_1,\dots,X_{s-1},Y_{s-1}$ is a sequence of distinct
necklaces representatives such that
\[X_i\in \sbinom{W^n}{m} \qquad Y_i\in \sbinom{W^n}{m+1},\]
and
\[X_i \subset Y_i \qquad Y_i \supset X_{i+1}.\]
If $\alpha^\ell X_0 \subset Y_{s-1}$ with $\gcd(\ell,\sbinom{n}{1}_q)=1$,
then
\[\cC=\cL,\alpha^{\ell}\cL,\alpha^{2\ell}\cL,\dots,\alpha^{\parenv{\sbinom{n}{1}_q-1}\ell}\cL,\]
is a cyclic $(n;q)$-subspace Gray code for the middle levels.
\end{theorem}
\begin{theorem}{\cite{Etz13}}
\label{th:etzcon}
For any $q$, a power of a prime, and $n=5$, there exists a sequence
as in Theorem \ref{th:etzneck}, resulting in a cyclic optimal
$(5;q)$-subspace Gray code for the middle levels.
\end{theorem}

While Theorem \ref{th:etzneck} refers to subspaces in the middle levels, it
can be easily generalized.
\begin{theorem}
\label{th:genneck}
Let $W^n$ be an $n$-dimensional vector space over $\gf(q)$, and let
$\alpha\in\gf(q)$ be a primitive element. Assume
$\cL=X_0,X_1,\dots,X_{s-1}$ is a path in $\cP_q(n)$ visiting only
representatives of distinct necklaces. If all the visited necklaces are
of equal size $N$, $\alpha^\ell X_0$ and $X_{s-1}$ are adjacent in
$\cP_q(n)$, and $\gcd(\ell,N)=1$, then
\[\cC=\cL,\alpha^{\ell}\cL,\alpha^{2\ell}\cL,\dots,\alpha^{\parenv{N-1}\ell}\cL,\]
is a cyclic $(n;q)$-subspace Gray code.
\end{theorem}
\begin{IEEEproof}
It can be easily verified that all adjacent elements in $\cC$ are
adjacent in $\cP_q(n)$ (including the first and last one), and since
all necklaces are of equal size, all the elements of $\cC$ are
distinct.
\end{IEEEproof}

We are now in a position to state and prove a construction for
$(5;q)$-subspace Gray codes.

\begin{theorem}
For any $q$, a power of a prime, there exists a cyclic optimal
$(5;q)$-subspace Gray code.
\end{theorem}

\begin{IEEEproof}
Let $W^5$ be a $5$-dimensional vector space over $\gf(q)$. Since $5$
is prime, the sizes of necklaces of dimensions $1$ through $4$ are all
the same and equal to $\sbinom{5}{1}_q$. In particular, this means
that there is exactly one necklace of dimension $1$, and exactly one
necklace of dimension $4$.

Let 
\[\cL=X_0,Y_0,X_1,Y_1,\dots,X_{s-1},Y_{s-1},\]
be a sequence of necklaces representatives, $\dim(X_i)=2$,
$\dim(Y_i)=3$, as in Theorem \ref{th:etzcon}, where
\[s=\left. \sbinom{5}{2}_q \right/ \sbinom{5}{1}_q=\left. \sbinom{5}{3}_q \right/ \sbinom{5}{1}_q=q^2+1\geq 2.\]

We construct $\cL'$ by reversing the order of $Y_0$ and $X_1$, and
inserting two new necklaces,
\[\cL'=X_0,X_0\cap X_1,X_1,Y_0,Y_0+Y_1,Y_1,\dots,X_{s-1},Y_{s-1}.
\]
Since $X_0,X_1\subset Y_0$, while $X_0\neq X_1$, $\dim(X_0)=\dim(X_1)=2$, and
$\dim(Y_0)=3$, we must have
\[\dim(X_0\cap X_1)=1.\]
Furthermore, $Y_0,Y_1\supset X_1$, and $Y_0\neq Y_1$, hence
\[\dim(Y_0+Y_1)=\dim(Y_0)+\dim(Y_1)-\dim(Y_0\cap Y_1)=4.\]

The sequence $\cL'$ clearly satisfies the requirements of Theorem
\ref{th:genneck}. Let $\cC'$ be the cyclic $(5;q)$-subspace Gray code
constructed in Theorem \ref{th:genneck} using $\cL'$. It is easily
seen that $\cC'$ contains all of the subspaces of $W^5$ except for
$W^5$ and $W^0$, the trivial $0$-dimensional subspace. We use a series
of sub-sequence reversals, similar to the above reversal, to make room
to insert $W^0$ and $W^5$.

The code $\cC'$ is comprised of sub-sequence blocks of the form
$\alpha^{i\ell}\cL'$,
\[\cC'=\cL',\alpha^\ell\cL',\alpha^{2\ell}\cL',\dots,\alpha^{\parenv{\sbinom{5}{1}_q-1}\ell}\cL'.\]
There are $\sbinom{5}{1}_q$ such blocks, each of length
$s+2=q^2+3$. We now zoom in on the first two blocks, $\cL'$ and
$\alpha^\ell \cL'$. First, in the block $\cL'$, we reverse the order
of the $3$rd, $4$th, and $5$th elements, thus obtaining
\[
\cL'' = X_0,X_0\cap X_1,Y_0+Y_1,Y_0,X_1,Y_1,\dots,X_{s-1},Y_{s-1}.
\]
We do the same in $\alpha^\ell\cL'$ and obtain $\alpha^\ell\cL''$. We
note that except for $X_0\cap X_1$ and $Y_0+Y_1$, any two adjacent
elements in the sequence are also adjacent in $\cP_q(n)$.

Next, in the combined two blocks $\cL'',\alpha^{\ell}\cL''$, we
reverse the sequence of elements starting from $Y_0+Y_1$ and ending
with $\alpha^{\ell}(X_0\cap X_1)$, and then insert $W^5$ and $W^0$ to
obtain
\begin{align*}
\cL^* &= X_0, X_0\cap X_1, W^0, \alpha^{\ell}(X_0\cap X_1), \alpha^{\ell}X_0, \\
& \quad\ Y_{s-1},X_{s-1}, Y_{s-2},X_{s-2} \dots,Y_2,X_2,\\
& \quad\ Y_1,X_1,Y_0,Y_0+Y_1, W^5, \alpha^{\ell}(Y_0+Y_1),\alpha^{\ell}Y_0,\\
& \quad\ \alpha^{\ell}X_1,\alpha^{\ell}Y_1,\dots,\alpha^{\ell}X_{s-1},\alpha^{\ell}Y_{s-1}.
\end{align*}
It is now easy to verify that $\cL^*$ describes a path in $\cP_q(n)$,
and that replacing the first two blocks in $\cC'$ with $\cL^*$ gives
\[\cC=\cL^*,\alpha^{2\ell}\cL',\alpha^{3\ell}\cL',\dots,\alpha^{\parenv{\sbinom{5}{1}_q-1}\ell}\cL',\]
which is indeed a cyclic optimal $(5;q)$-subspace Gray code.
\end{IEEEproof}

We remark in passing that the choices for which sub-sequences to
reverse in the proof, were made specific for ease of presentation. A
similar more general construction can be described, in which the
reversal process allows for more choices of reversal positions.

\section{Conclusion}
\label{sec:conc}

We studied optimal Gray codes for subspaces in two settings: the
Grassmann graph, and the projective-space graph. In the first case we
were able to construct cyclic optimal Gray codes for all parameters
using a recursive construction. In addition, simple recursive encoding
and decoding functions were provided. These algorithm induce an
enumerative-coding scheme, which is at least as efficient as known
schemes, and for certain parameters, surpasses them.

In the case of the projective-space graph, it was shown that there are
no optimal Gray codes (cyclic or not) in the projective-space graph of
even dimension. For odd dimensions, we were able to show a
construction for dimensions $3$ and $5$, which are derived from
constructions for the middle-levels problem of the same dimension.

Two related open questions arise: the first is whether there exist
cyclic optimal subspace Gray codes for \emph{all} even dimensions. The
second questions, is whether a reverse connection exists which derives
optimal codes for the middle-levels problem from a subspace Gray
code. Even in $3$ dimensions the answer to the latter is not clear.

\section*{Acknowledgments}

The author would like to thank Tuvi Etzion for valuable discussions
which contributed to the content of this paper. The author would also
like to thank Muriel M\'edard for hosting him at MIT during his
sabbatical.

\bibliographystyle{IEEEtranS}
\bibliography{allbib}

\end{document}